\documentclass[manuscript]{aastex}

\usepackage{multirow}
\usepackage{enumerate}

\shorttitle{A Survey of Astronomical Research}
\shortauthors{Ribeiro, Russo, \& C\'ardenas-Avenda\~no}

\begin{document}


\title{A Survey of Astronomical Research: \\ An Astronomy for Development Baseline}


\author{V. A. R. M. Ribeiro\altaffilmark{1}}
\affil{Astrophysics, Cosmology and Gravity Centre, Department of Astronomy, University of Cape Town, Private Bag X3, Rondebosch 7701, South Africa.}
\email{vribeiro@ast.uct.ac.za}

\author{P. Russo}
\affil{EU Universe Awareness, Leiden Observatory, Leiden University, PO 9513 Leiden, 2300 RA, the Netherlands.}
\email{russo@strw.leidenuniv.nl}

\and

\author{A. C\'ardenas-Avenda\~no}
\affil{Departamento de F\'isica, Universidad Nacional de Colombia, Carrera 45 No 26 -- 85, Edificio Gutierr\'ez, Bogot\'a, DC Colombia.}


\altaffiltext{1}{South African Square Kilometer Array Fellow.}


\begin{abstract}
Measuring scientific development is a difficult task. Different metrics have been put forward to evaluate scientific development; in this paper we explore a metric that uses the number of peer-reviewed, and when available non-peer-reviewed articles, research research articles as an indicator of development in the field of astronomy. We analyzed the available publication record, using the SAO/NASA Astrophysics Database System, by country affiliation in the time span between 1950 and 2011 for countries with a Gross National Income of less than 14,365 USD in 2010. This represents 149 countries. We propose that this metric identifies countries in `astronomy development' with a culture of research publishing. We also propose that for a country to develop astronomy it should invest in outside expert visits, send their staff abroad to study and establish a culture of scientific publishing. Furthermore, we propose that this paper may be used as a baseline to measure the success of major international projects, such as the International Year of Astronomy 2009.
\end{abstract}





\section{Introduction}\label{intro}
Astronomy is a fascinating subject, with an unique ability to inspire and to stimulate curiosity in human beings about the wonders of science and technology. This makes astronomy useful tool for bringing science to the general public, to inspire, to show the scientific method and to open their eyes to a new prespective. Astronomy has shown to have wide-reaching applications in many different sectors of society. One immediate example are the technological developments that came from the building of the ESA/NASA {\it Hubble Space Telescope}\footnote{\url{spinoff.nasa.gov} last accessed 2012 November 02}. Such as, the use of mirror technology to increase semiconductor productivity and performance, and CCD technology being adapted for more efficient biopsies\footnote{\url{spinoff.nasa.gov/pdf/Hubble\_Flyer.pdf} last accessed 2012 November 02} \citep{N01b}. This demonstrates that astronomy not only aims to answer fundamental questions about how the Universe works and to stimulate curiosity, but can also aid technology development and economical growth. Although it is difficult to quantify the return of investment in astronomy, some reports show spin-off technology development to return as much as ten-to-one\footnote{\url{http://www.ic.gc.ca/eic/site/cprp-gepmc.nsf/vwapj/Coalition\_Canadian\_Astronomy.pdf/\%24FILE/Coalition\_Canadian\_Astronomy.pdf} last accessed 2012 November 02}.

One project that aimed to stimulate and inspire people's curiosity with the wonders of the Universe was the International Year of Astronomy in 2009 (IYA2009). This project reached over 815 million people in 148 countries \citep{RC10} through various activities from star parties and school programs to the use of IYA2009 to launch of University programs \citep[e.g.][]{RPB11}. The success of IYA2009 was no mean feat. However, truly understanding and evaluating its impact, at least in astronomy, will be a hardeous task. Project evaluations are applied for numerous reasons\footnote{\url{http://www.astronomy2009.org/static/resources/iya2009\_evaluation\_guide\_spocs.pdf} last accessed 2012 July 20}, for example: (i) to determine if the project goals were reached, (ii) to obtain information on the outcomes of an event, along with suggestions for improvement, (iii) to identify the changes resulting from the implementation of a project, (iv) to identify ways in which the project could have been more effective and efficient, (v) to identify unexpected results, (vi) to crystallize ideas about the event and what it is intended to achieve, (vii) to provide encouragement by demonstrating that efforts have been worthwhile. Measuring this impact may be done in various forms.
	
Developing Astronomy Globally (DAG), a cornerstone project of IYA2009 which was fed into the International Astronomical Union (IAU) Strategic Plan\footnote{\url{http://www.iau.org/static/education/strategicplan\_091001.pdf} last accessed 2012 July 20}, was designed to develop astronomy professionally (at universities and research level) worldwide\footnote{Developing Astronomy Globally, \url{http://www.developingastronomy.org/} last accessed 2012 July 20 }. As part of DAG, a survey was conducted as a self-evaluation of the countries participating in IYA2009 \citep{NG09}. The survey was completed by the IYA2009 Single Point of Contact from each country and therefore may suffer some bias and data may be incomplete. \citet{NG09} proposed that each country would fall into one of four separate phases of astronomy development, and presented some recommendations for development accordingly. In summary, these phases were: (i) well established, (ii) in need of support, (iii) non-existent with strong potential and (iv) non-existent with limited potential.

\citet{H07} extracted statistical information from the SAO/NASA Astrophysics Data System (hereafter ADS) to obtain an overview of the state of astronomy development per country. \citeauthor{H07} found that the number of publications per IAU member correlates strongly with Gross Domestic Product (GDP) per capita. However, this concentrated only on IAU member states and a select few non-member countries.
	
	This paper looks at a number of countries, most not included in \citet{H07}, sample of peer-reviewed research articles to measure countries in `Astronomy Development', and to identify those with a culture of publishing, using the ADS which is used by the entire astronomical community \citep{HKA09}, therefore a good database for use determining the astronomical research being carried out throughout the world. Other means are possible for this study, e.g. the World of Science. However, two major factors played a role in the decision to use ADS instead: (i) it is free and (ii) as mentioned above this is used by the entire astronomical community. From this viewpoint, counting the number of publications in astronomy by each country provides, to a first approximation, a good indicator of astronomy development.
	
	Due to the sheer amount of data and different sociological reasons for a country to be in `Astronomy Development' we only concentrate on providing a quantitative, rather than qualitative, discussion and invite the community to draw their own conclusions for their particular regions. In section~\ref{methods} we present the method used for these studies. Section~\ref{results} shows the results and in section~\ref{discussion} we discuss and draw conclusions from our findings.

\section{Methods}\label{methods}
When publishing in a refereed journal authors are required to provide their institution address with the article. In the majority of the cases this is also indexed for searching, alongside co-authors, title, and other key information that make searching for a journal article simple. We therefore used the ADS affiliation field to count astronomical publications by country from 1950 up to, and including, 2011. We only queried the Astronomy database for these studies. However, this query returns journals not only related to Astronomy but also from the Geosciences.

Particular care was taken for countries which may conflict with other words in the affiliation. For example, Niger is easily confused with Nigeria and Guinea is easily confused with Equatorial Guinea, Guinea-Bissau and Papa New Guinea. The search returned the number of papers for each country in a given year. We then selected papers, based on the biased view of the authors, that we consider main stream astronomy journals. These were: Astronomical Journal (including Supplements), Astronomy and Astrophysics, Astrophysical Journal (including Letters and Supplements), Monthly Notices of the Royal Astronomical Society, New Astronomy (including Reviews) and Physical Reviews. The first four journals were described by \citet{HKA09} as core journals read regularly by active astronomers.

	We searched, using the ADS Mighty Search\footnote{\url{http://adsabs.harvard.edu/mighty\_search.html}}, both refereed and non-refereed papers, although the latter are very difficult to quantify due to the fact that, in the majority of the cases, the affiliations are not given in the ADS abstract. Furthermore, counting the number of papers per country was based solely on whether the country name appeared in the affiliation field. As an example, the current paper each of the country in the affiliation field would receive a count of one paper. This is not uncommon in astronomy, where 55\% of papers has been suggested to arise from authors from different countries \citep{A07}.

	The number of papers published per year was used to identify which countries are in `Astronomy Development'. The selection of countries we considered based on their Gross National Income (GNI)\footnote{\url{http://data.un.org/} last accessed 2012 October 2012}. We considered those countries which have a GNI of less than 14,365 USD (based on the average world GNI for 2010). We should note that a country's GNI can be very dynamic. However, for the purpose of these studies just considering the 2010 is sufficient for the majority of the world's countries. This search retrieved 149 countries (Appendix~\ref{wc}), including all the Least Developed Countries (LDCs, Appendix~\ref{ldc}).

\section{Results}\label{results}
\begin{figure}
\centering
\plottwo{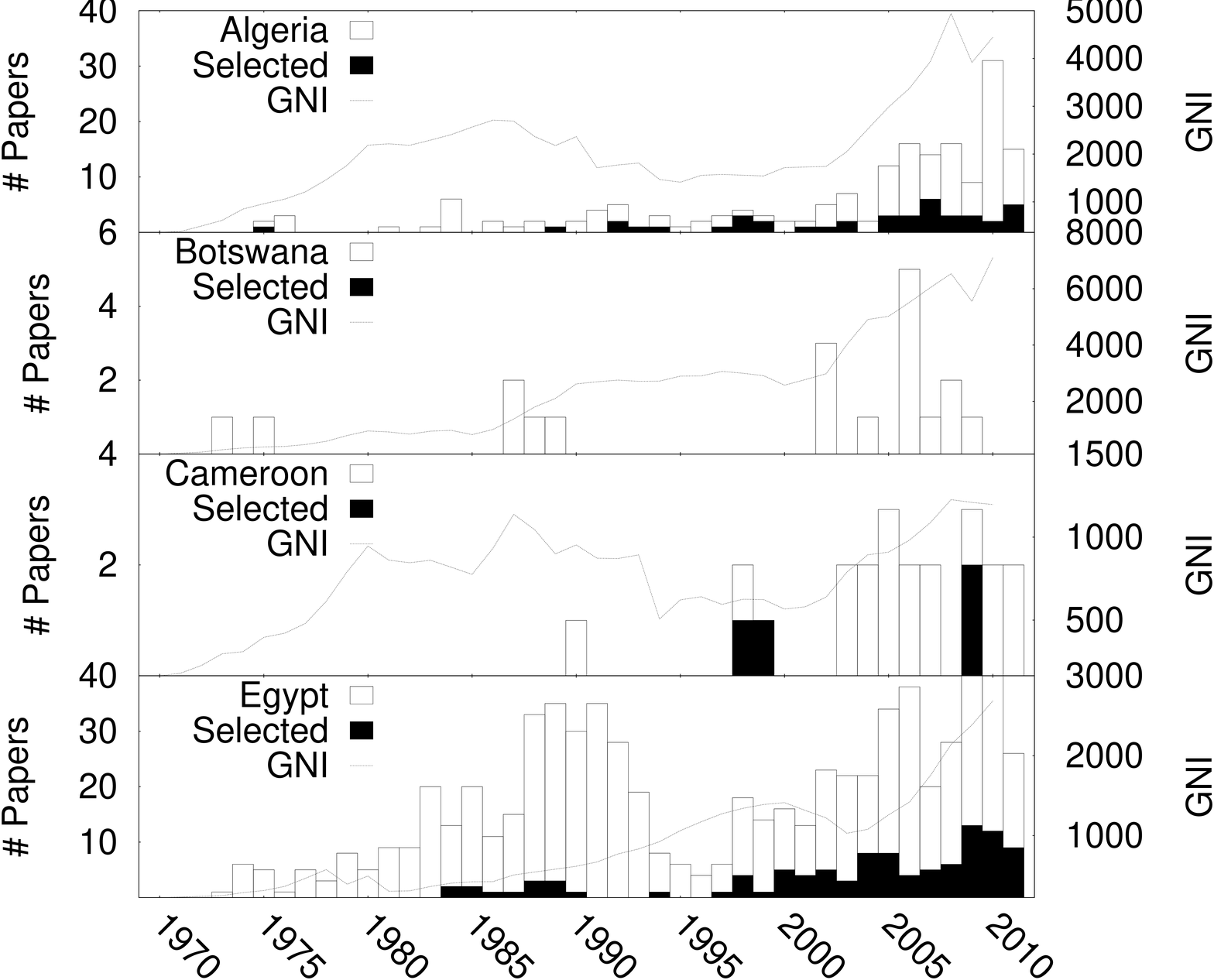}{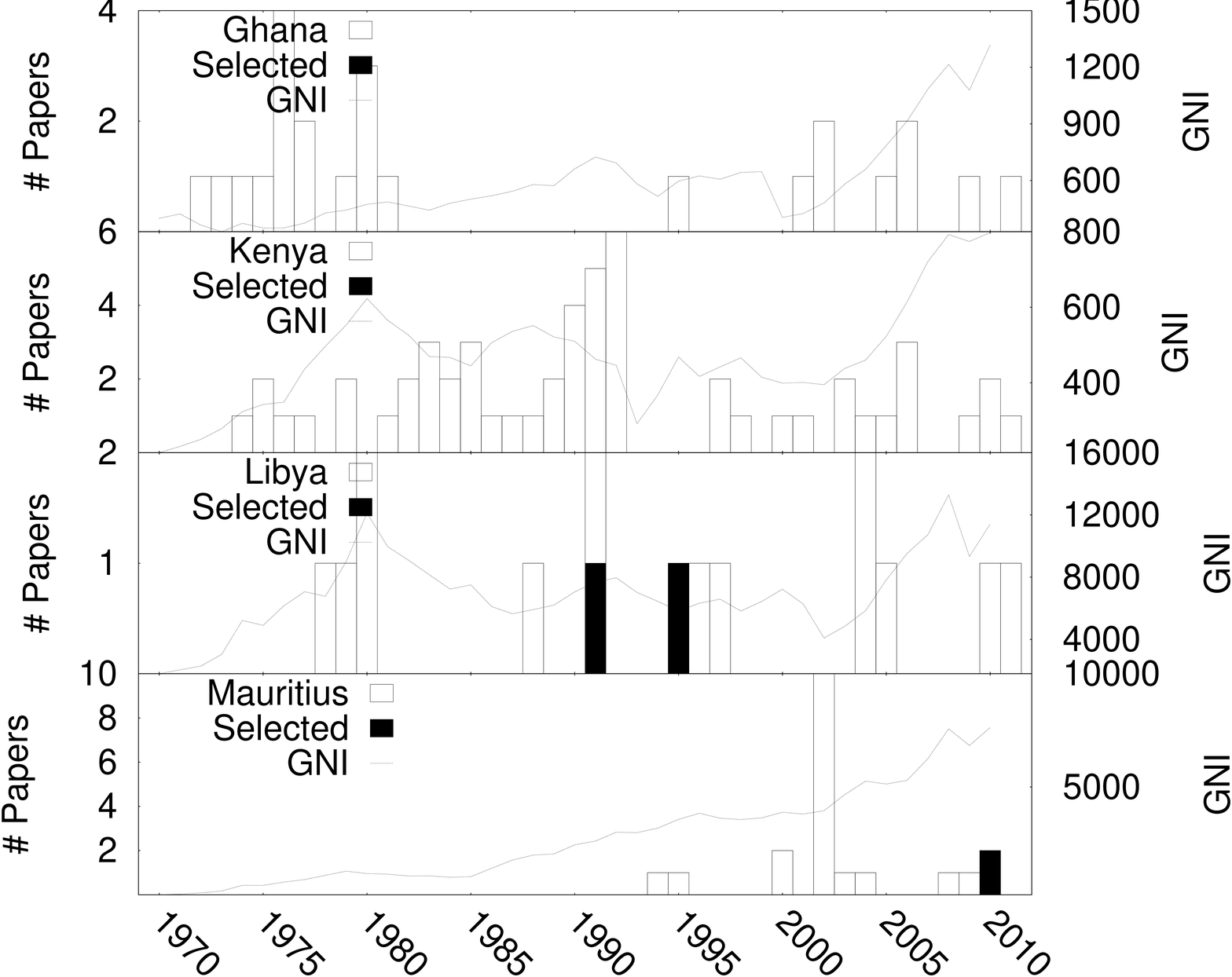}
\plottwo{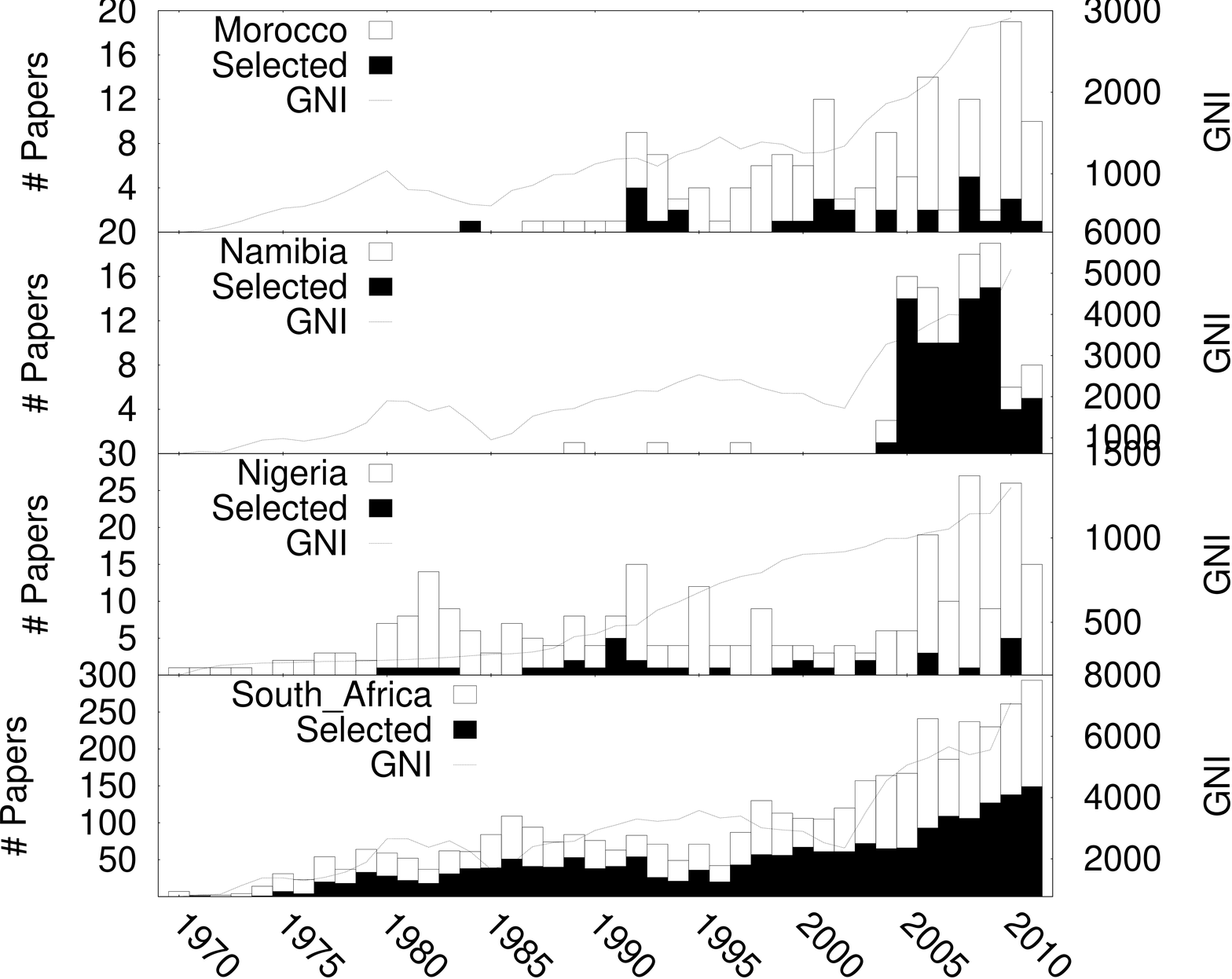}{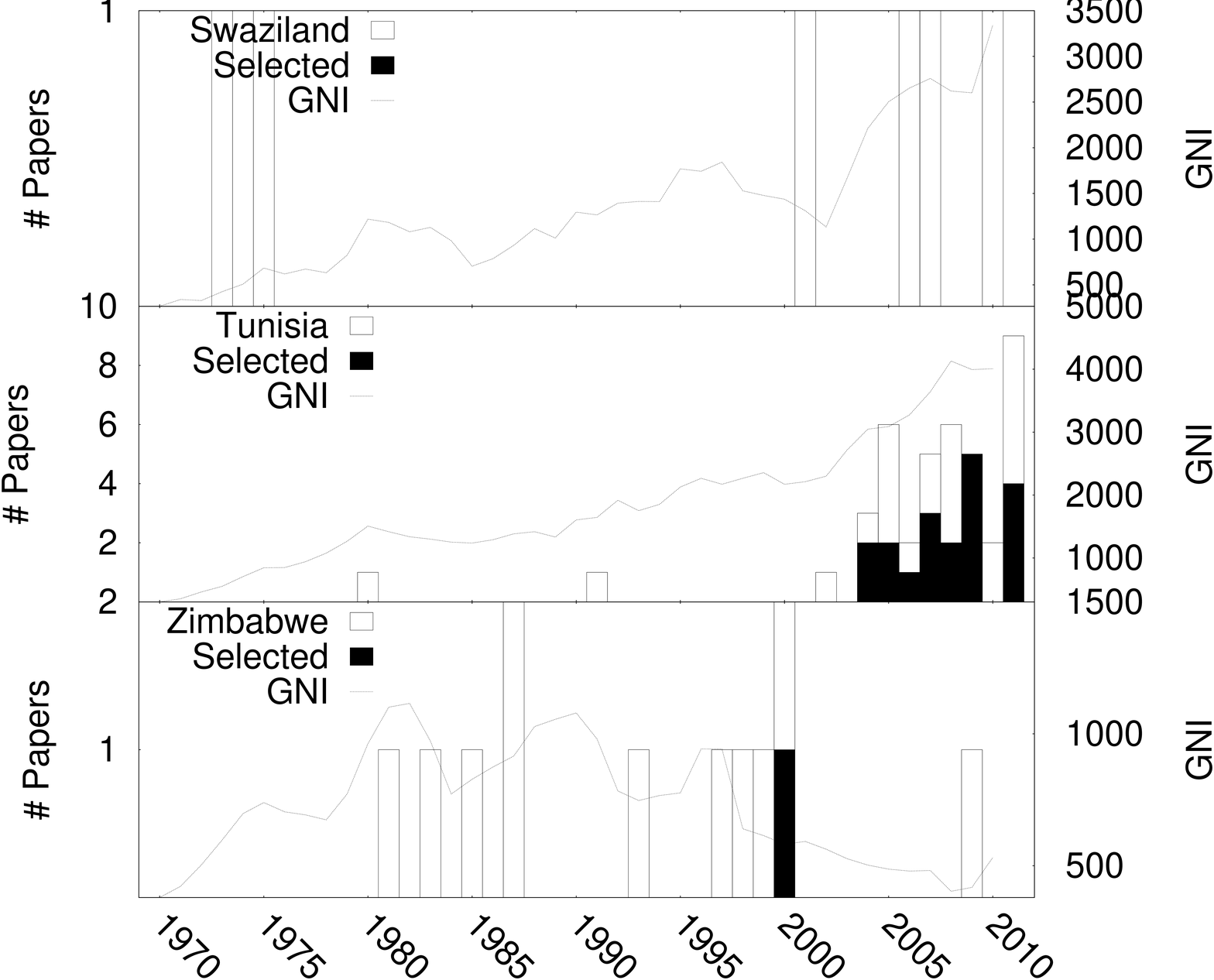}
\caption{Results for the African continent. Shown are all the results from the ADS search (white histograms) along with the journals identified as main streams (black histograms) and for comparison the countries GNI (dashed line).}
\label{fig1}
\end{figure}

\begin{figure}
\centering
\plottwo{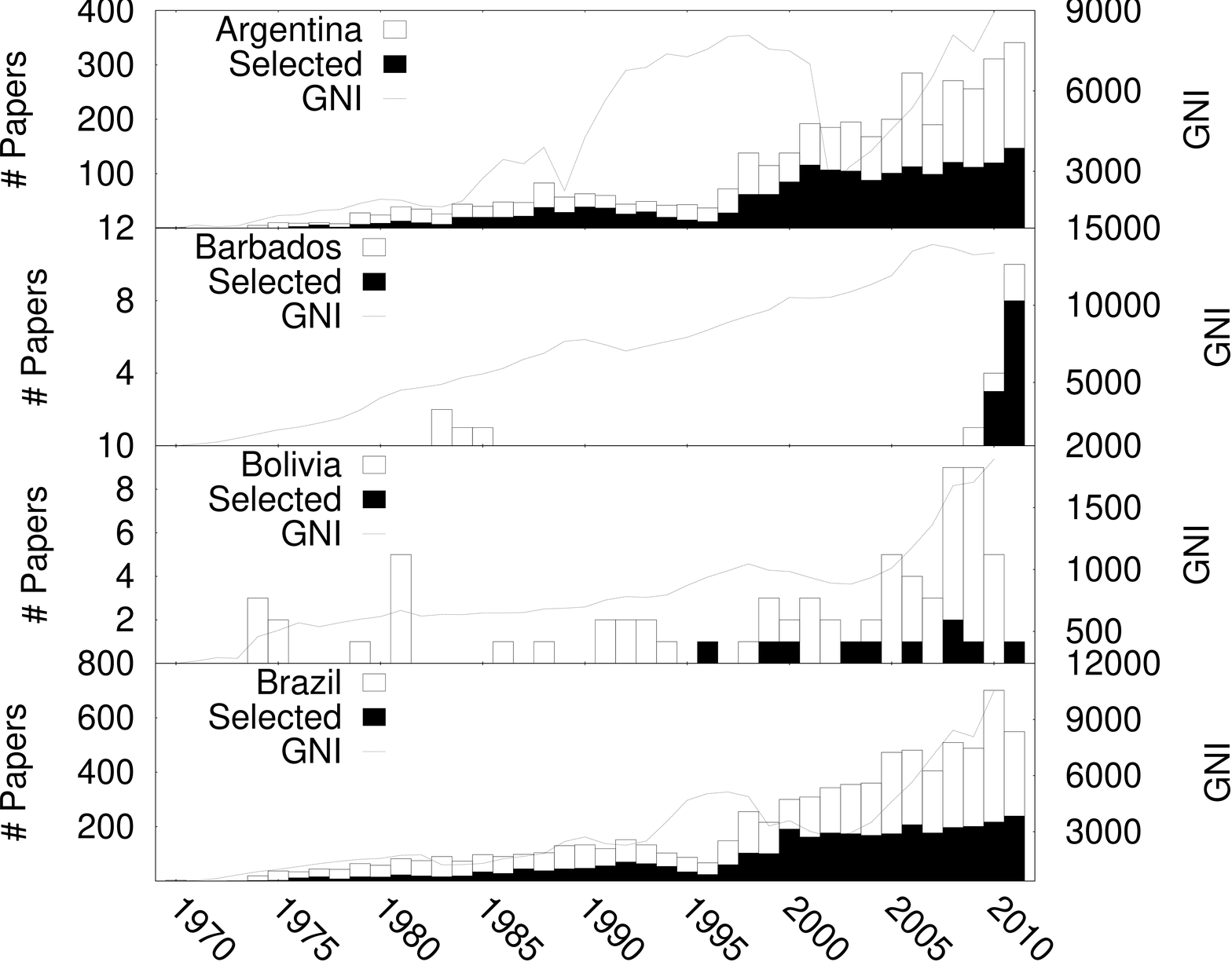}{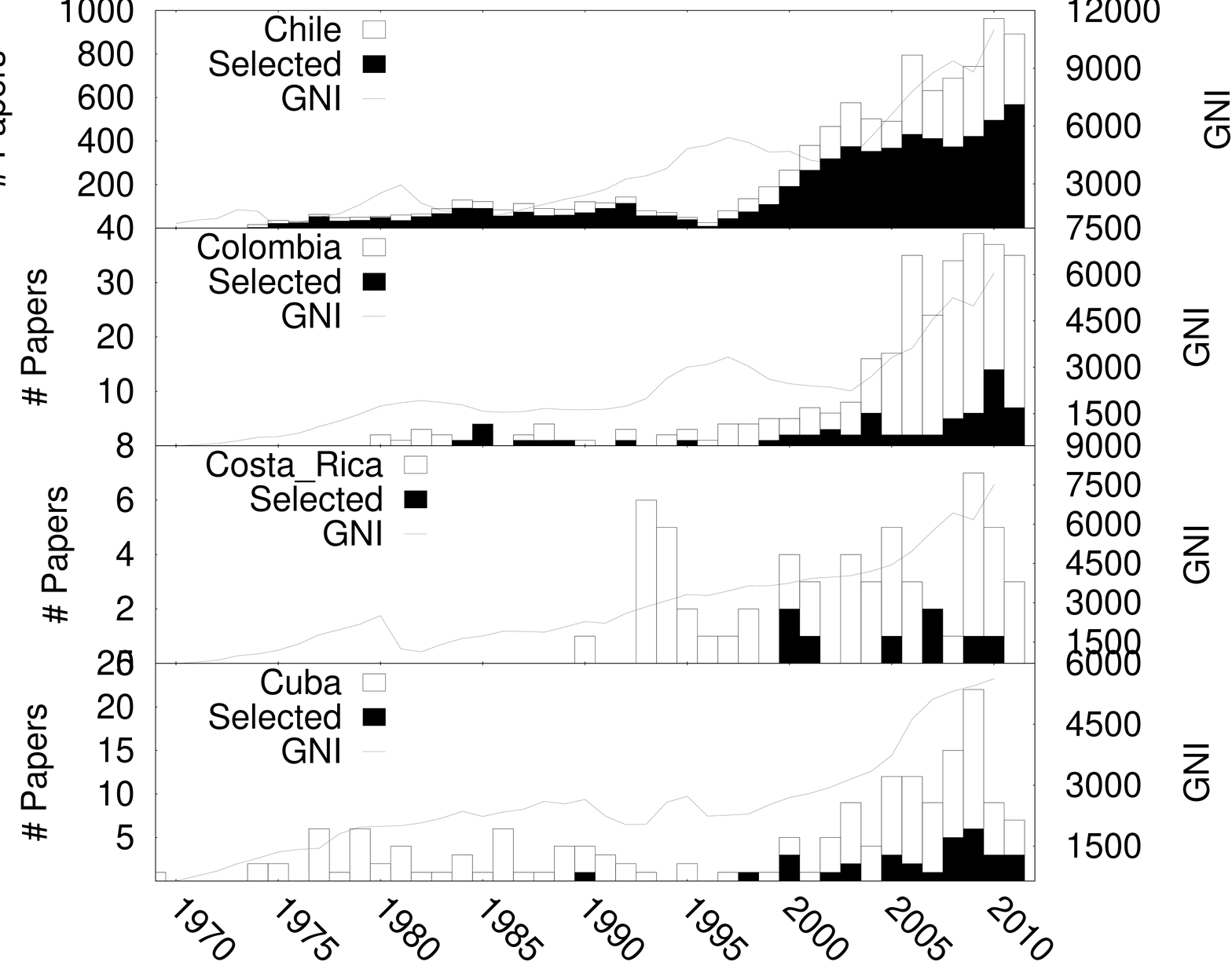}
\plottwo{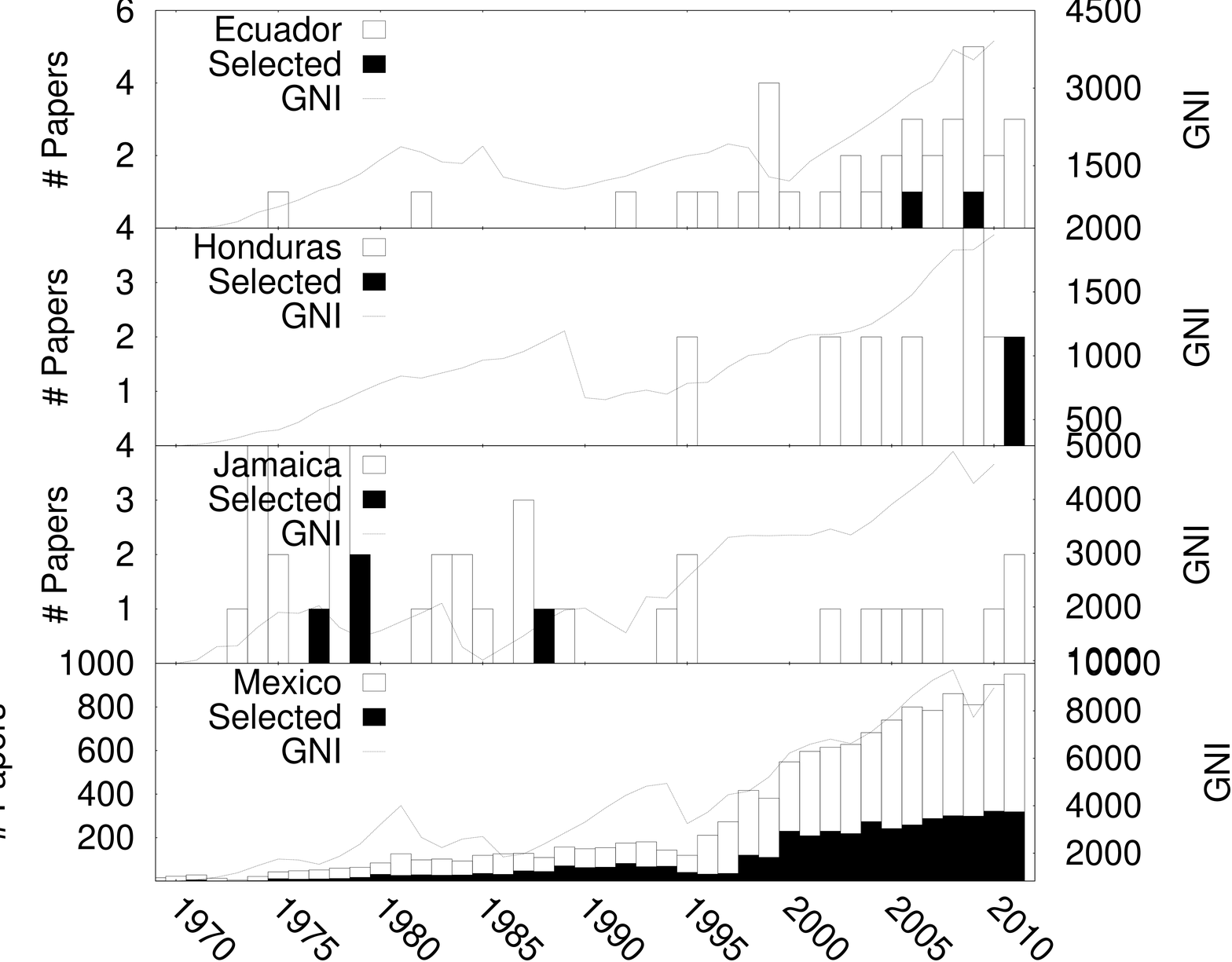}{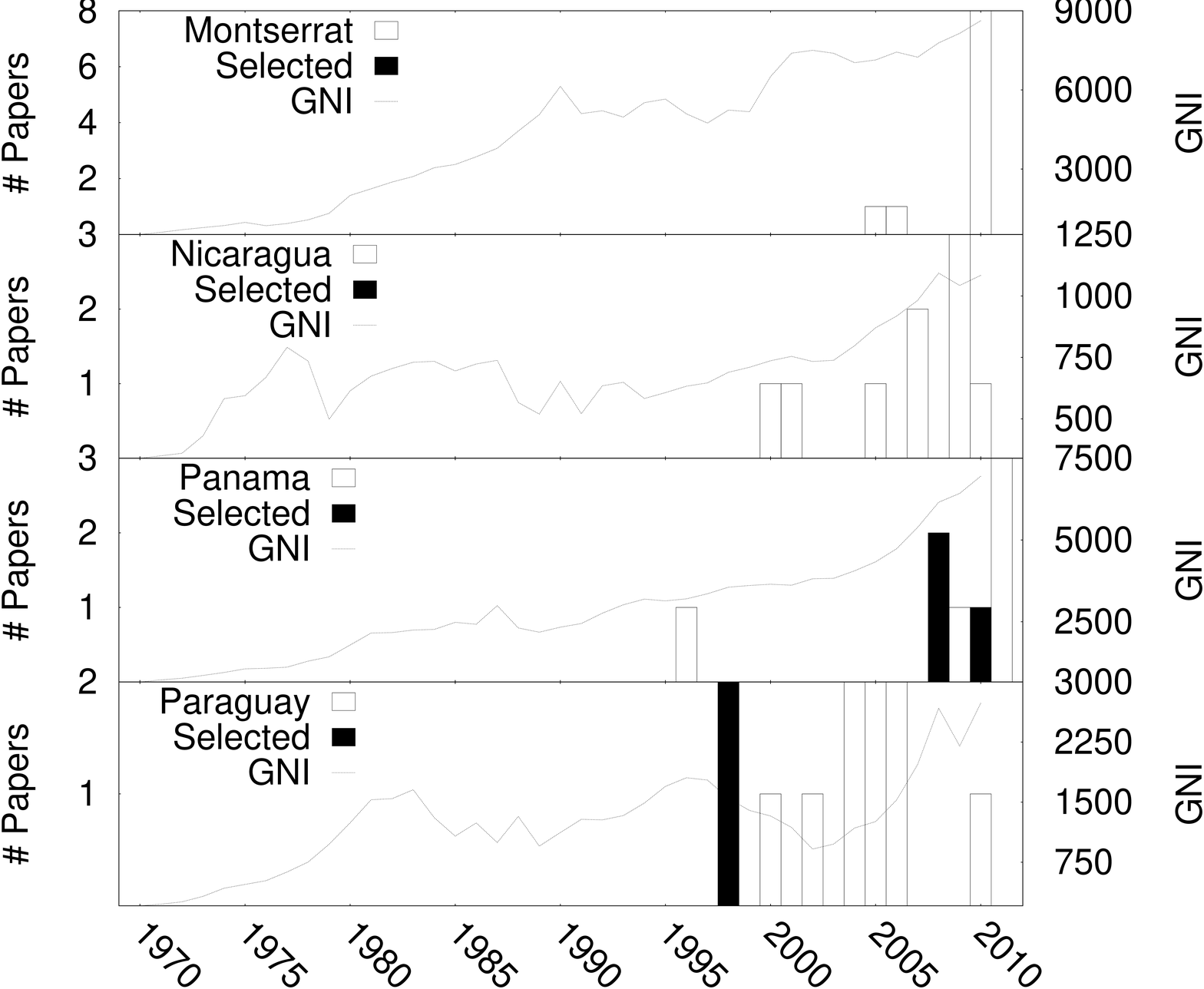}
\plotone{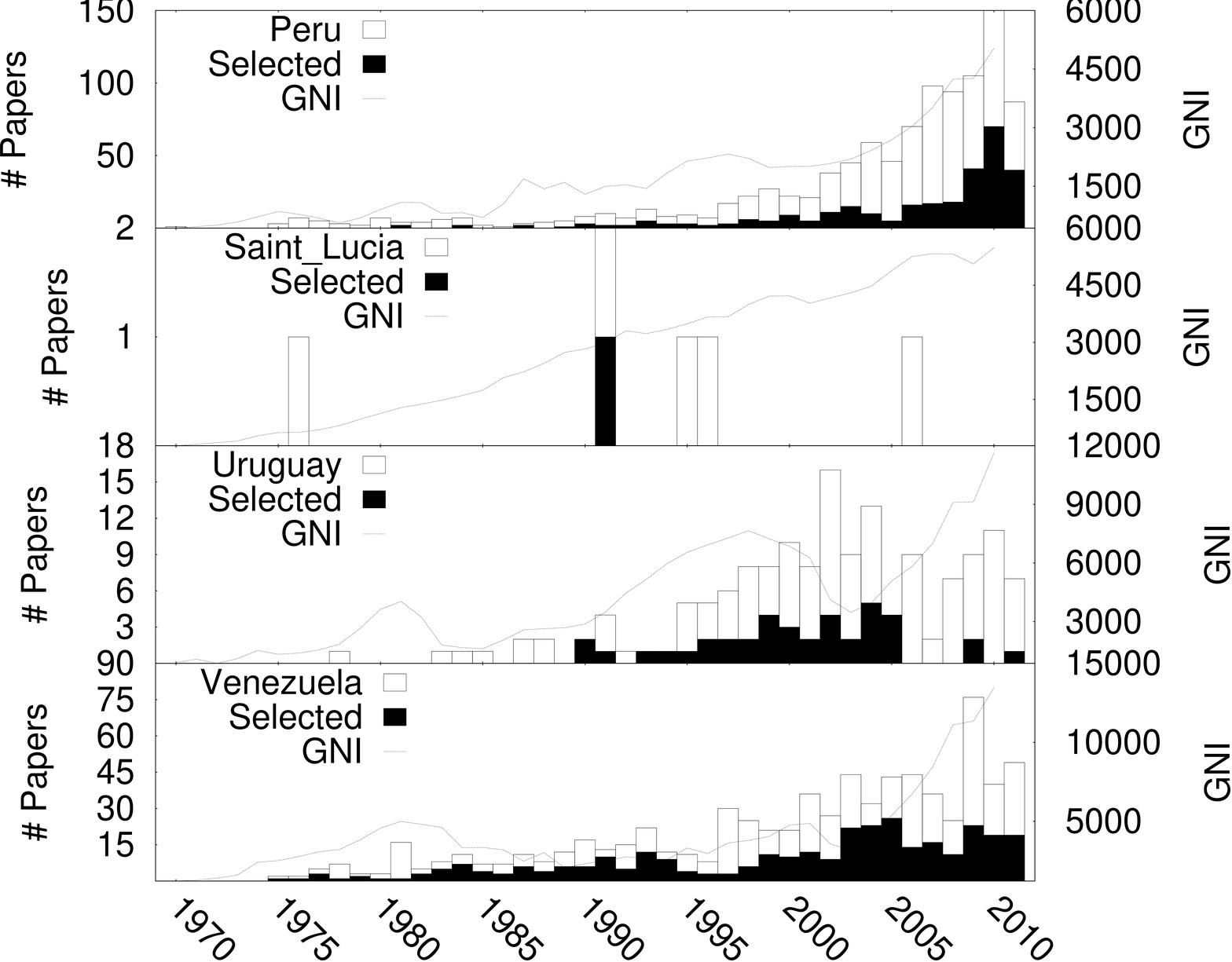}
\caption{As Fig.~\ref{fig1} but for South and Latin America.}
\label{fig2}
\end{figure}

\begin{figure}
\centering
\plottwo{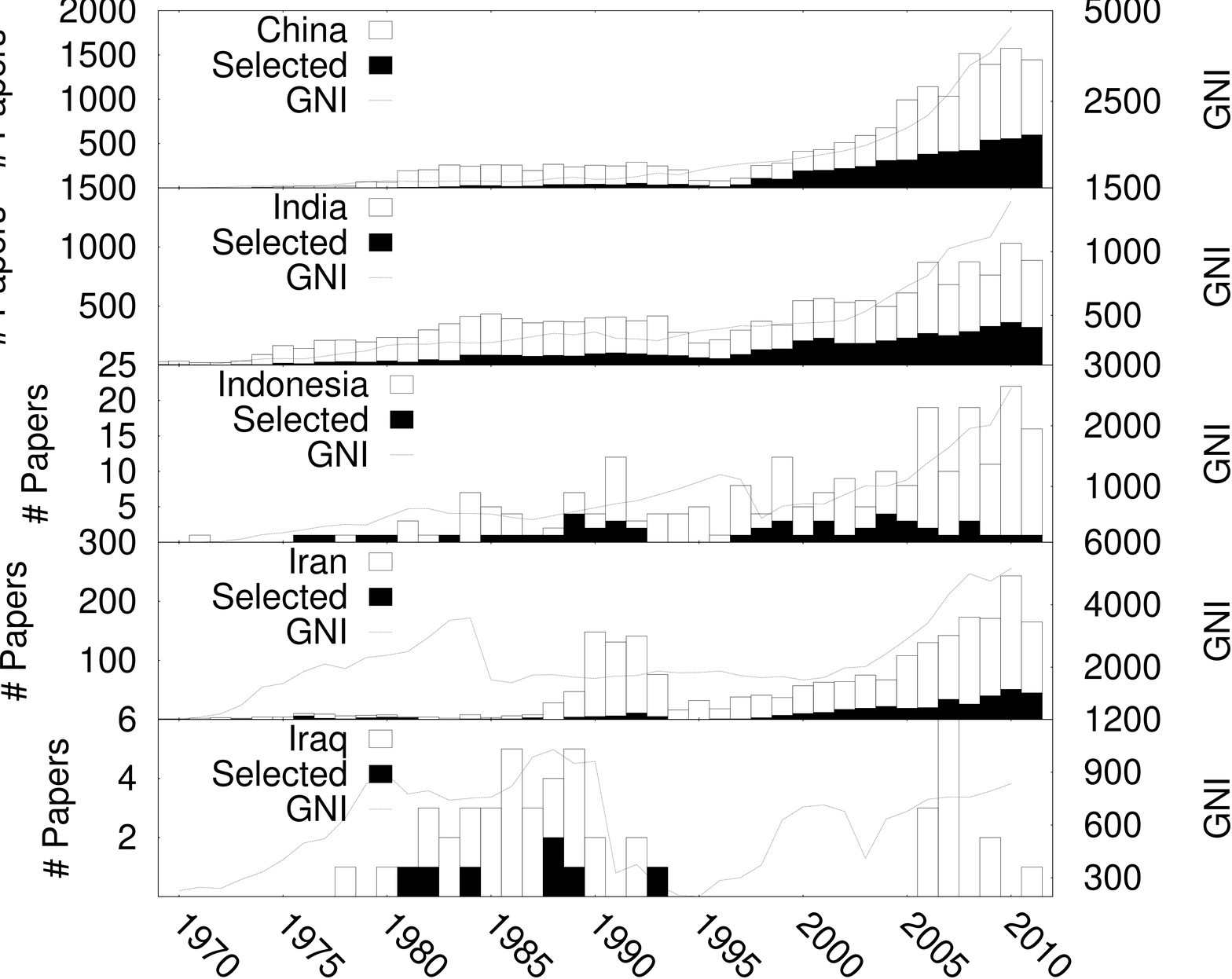}{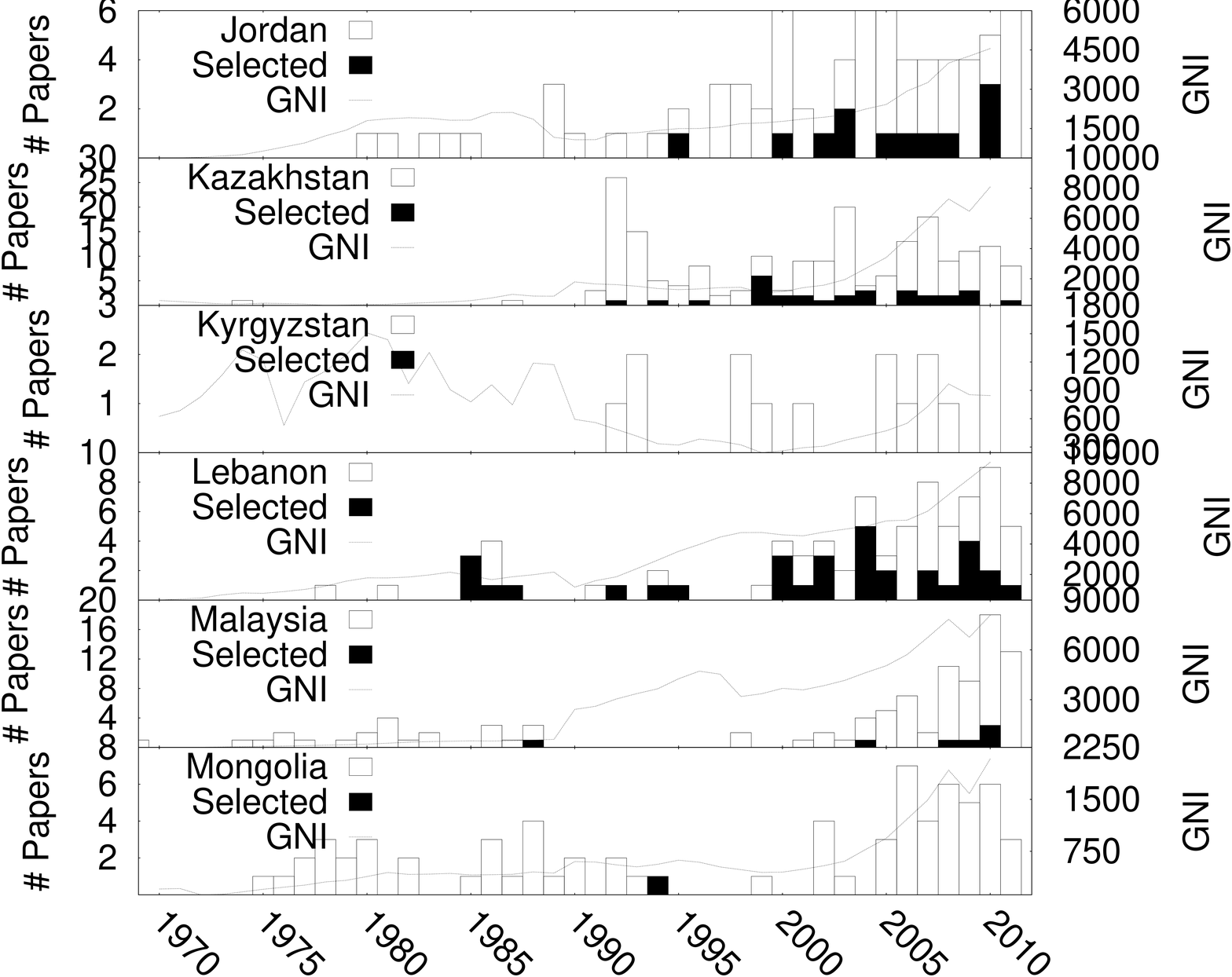}
\plottwo{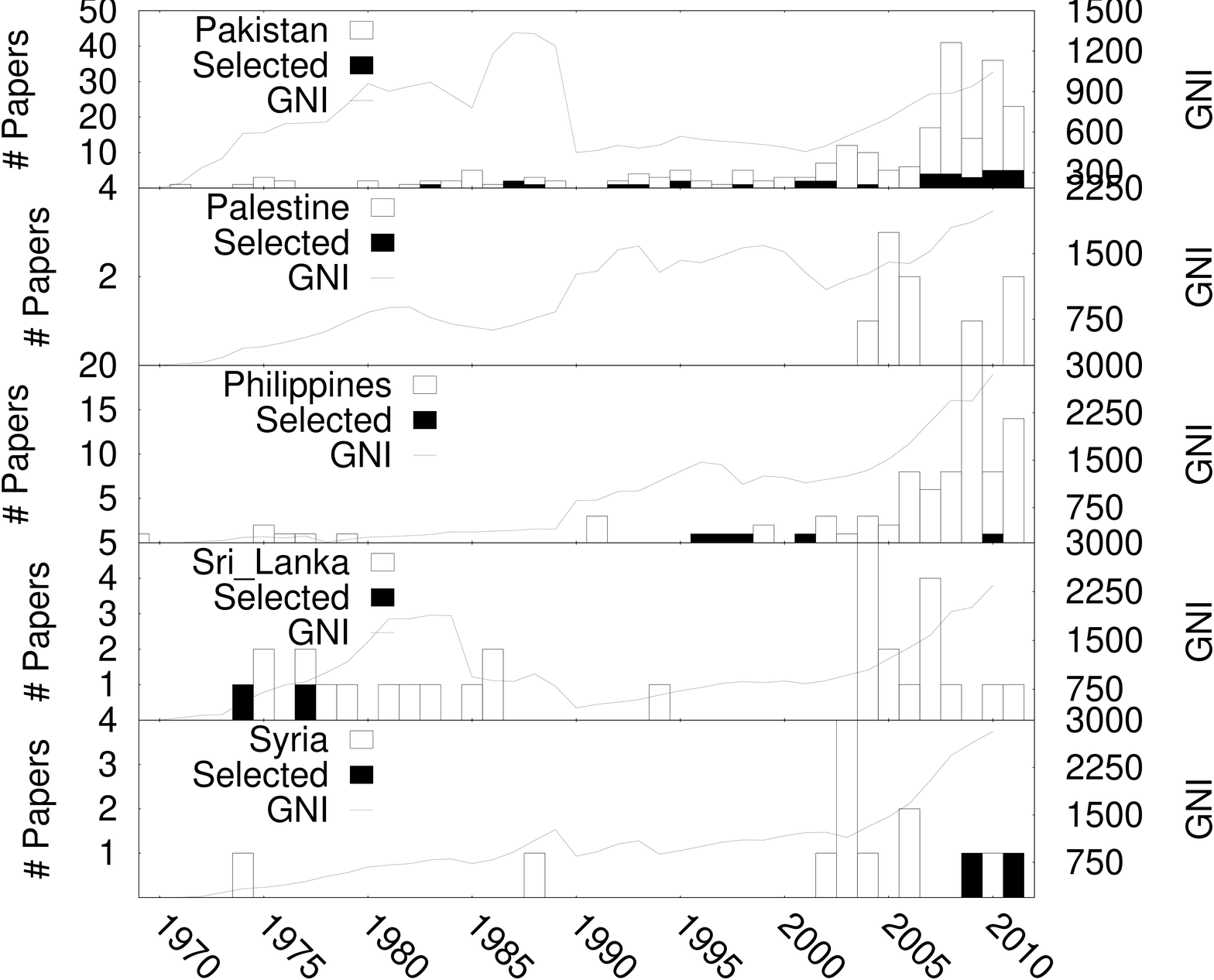}{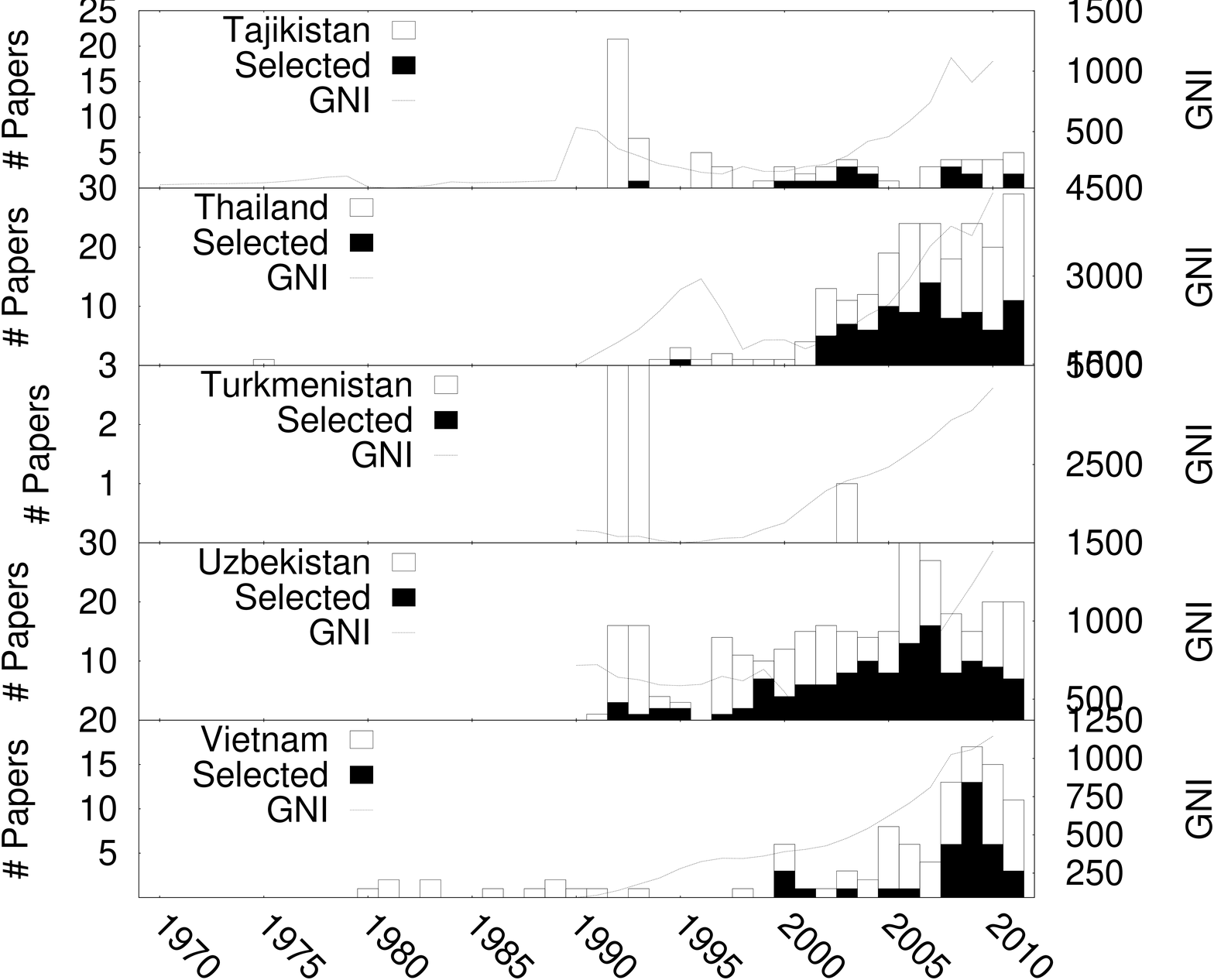}
\caption{As Fig.~\ref{fig1} but for Asia.}
\label{fig3}
\end{figure}

\begin{figure}
\centering
\plottwo{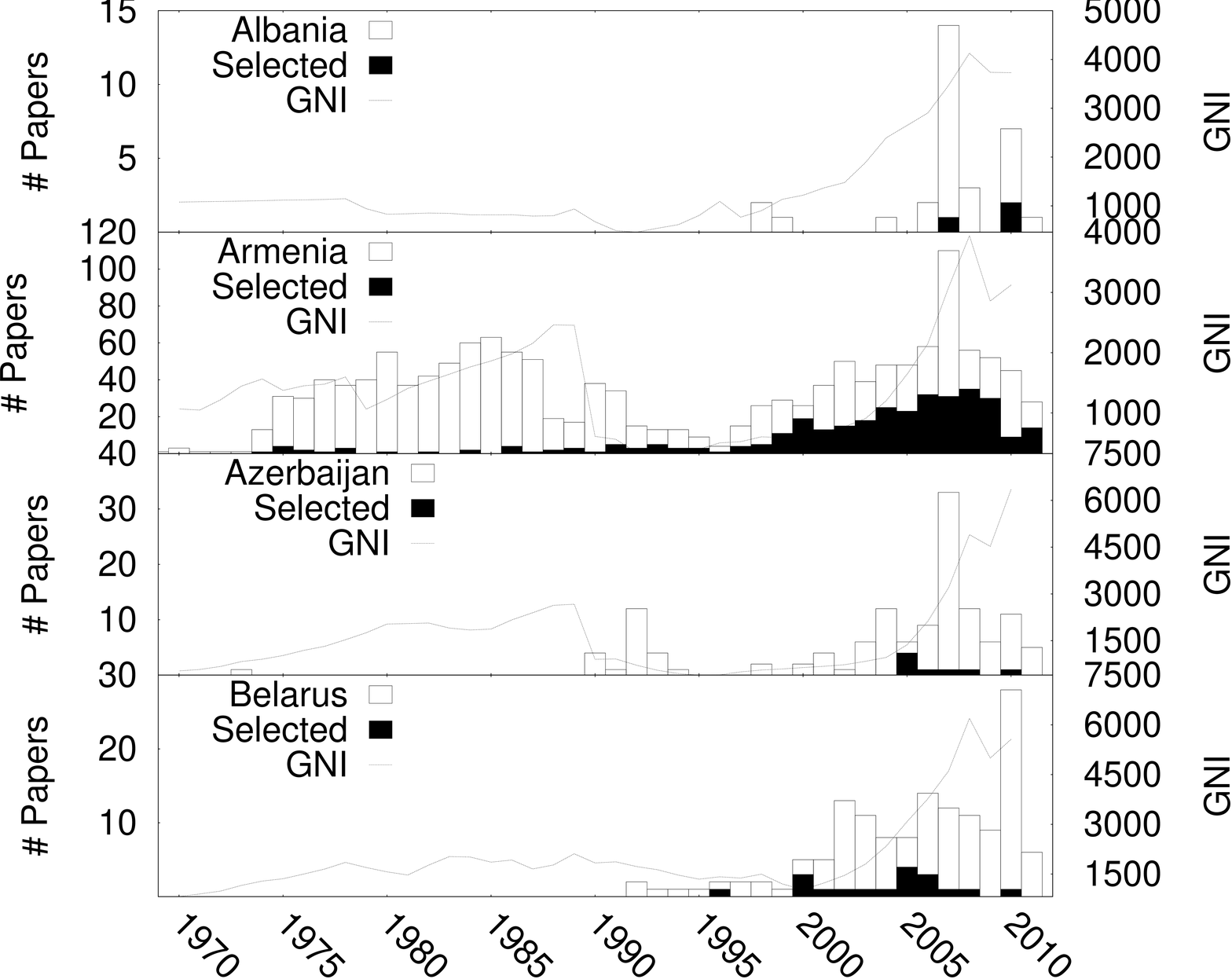}{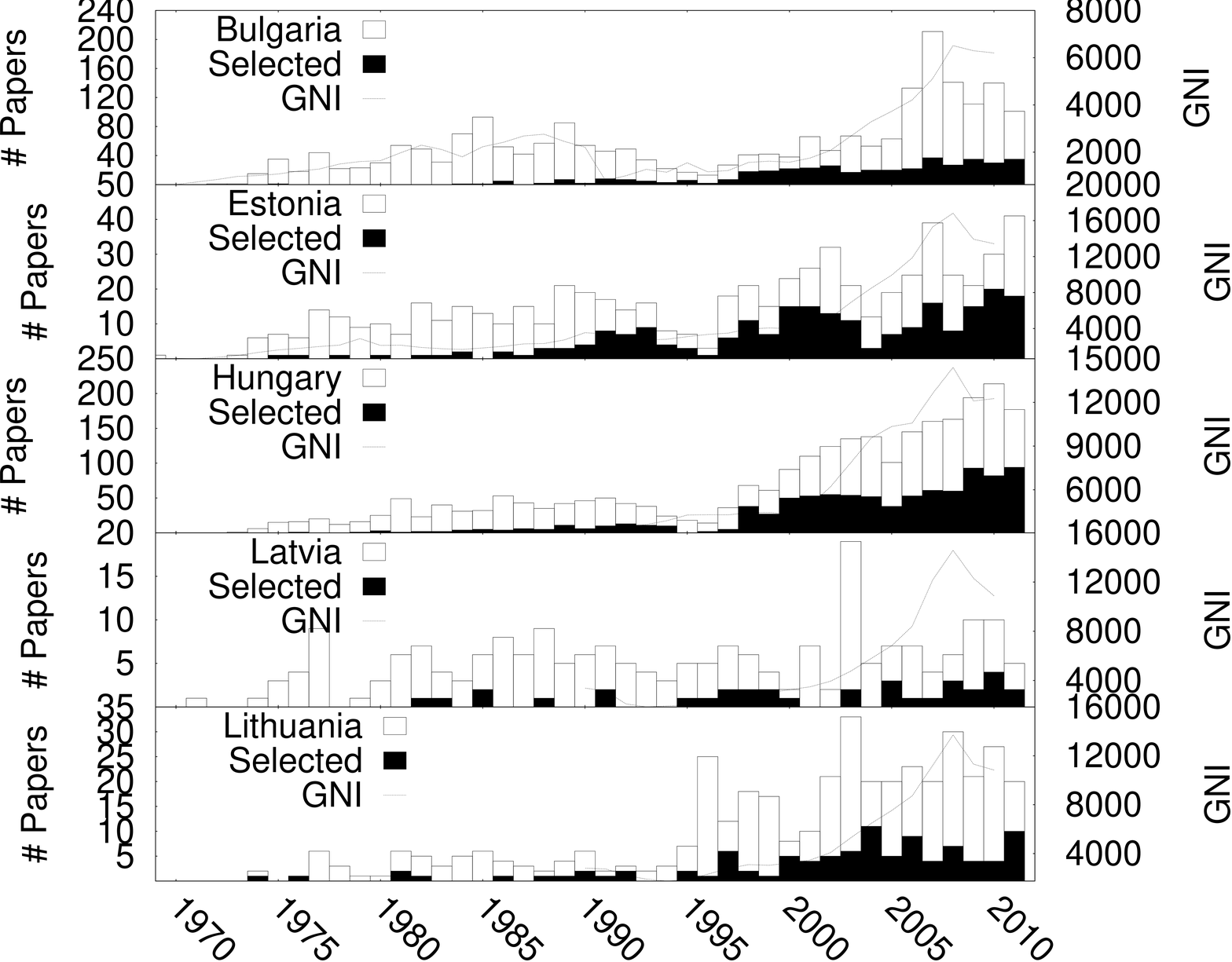}
\plottwo{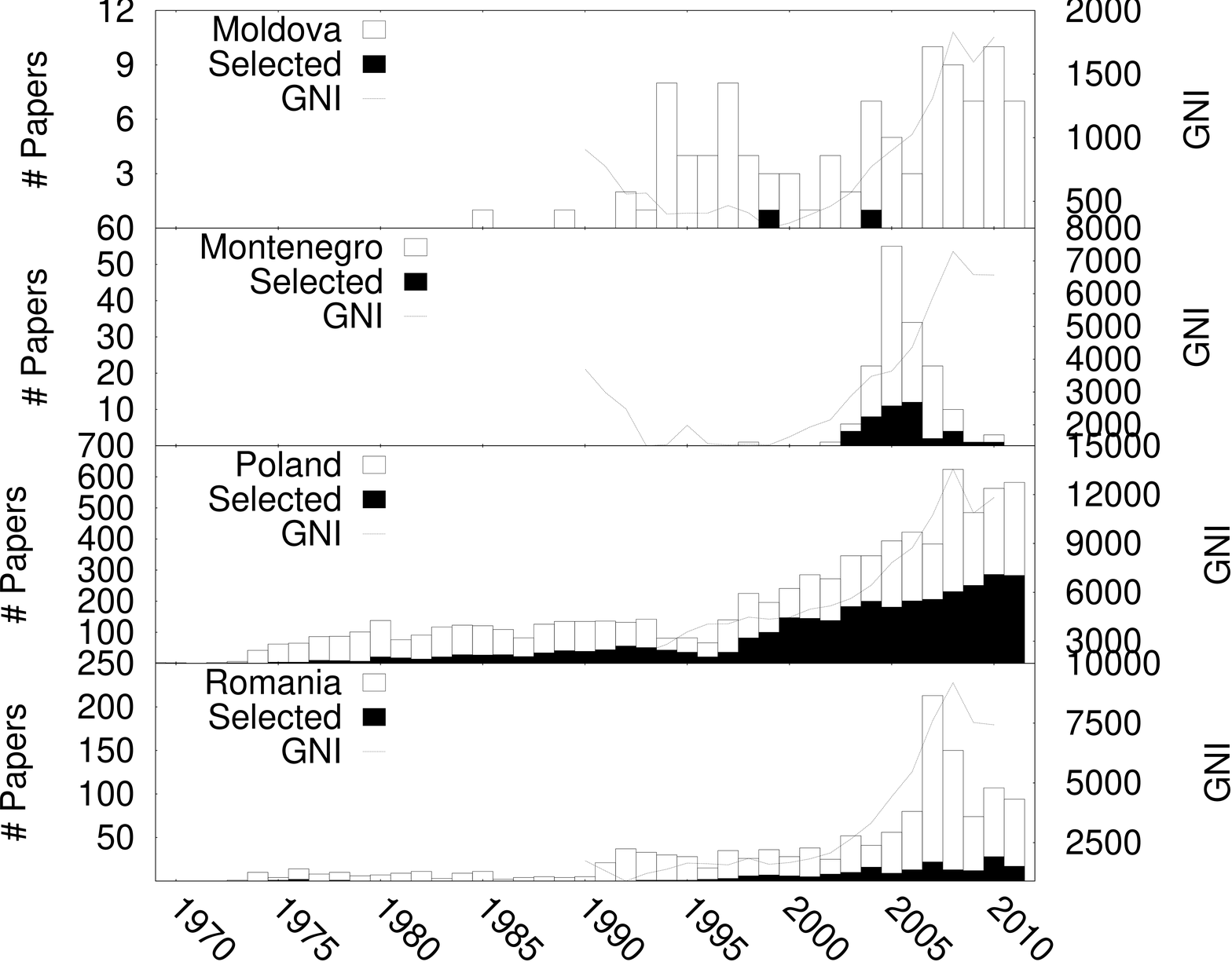}{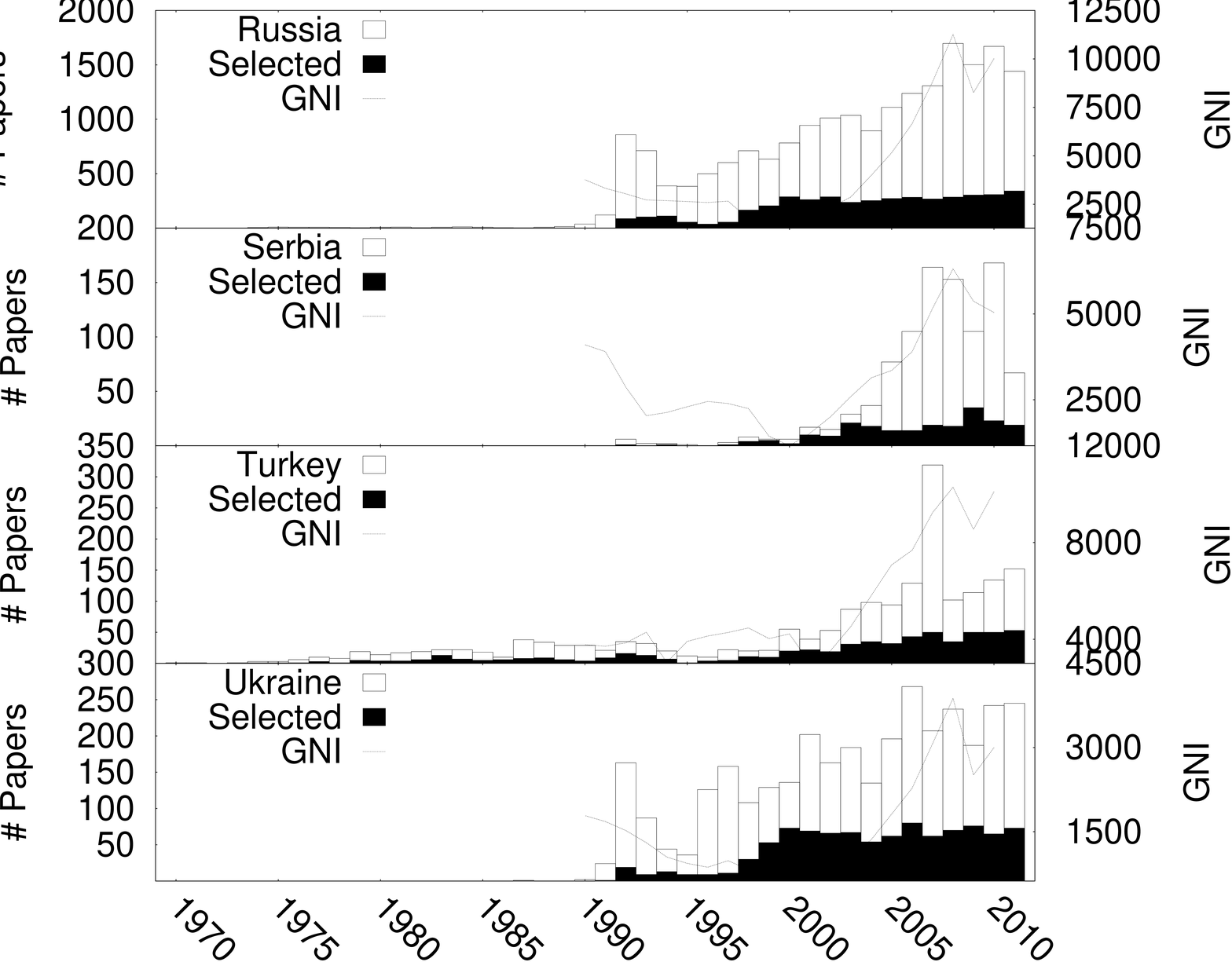}
\caption{As Fig.~\ref{fig1} but for Europe.}
\label{fig4}
\end{figure}

\begin{figure}
\centering
\plotone{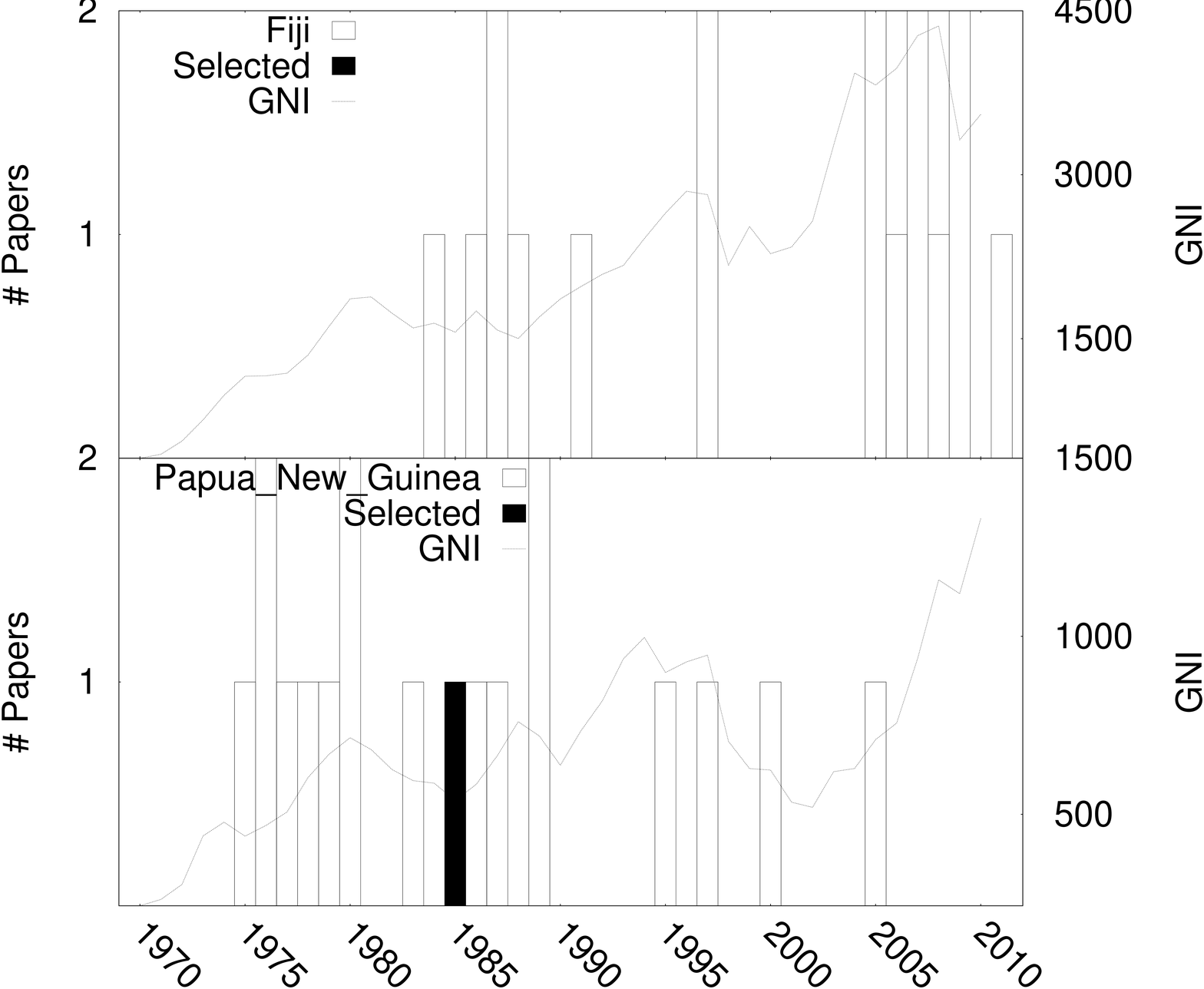}
\caption{As Fig.~\ref{fig1} but for Oceania.}
\label{fig5}
\end{figure}

\begin{figure}
\centering
\plottwo{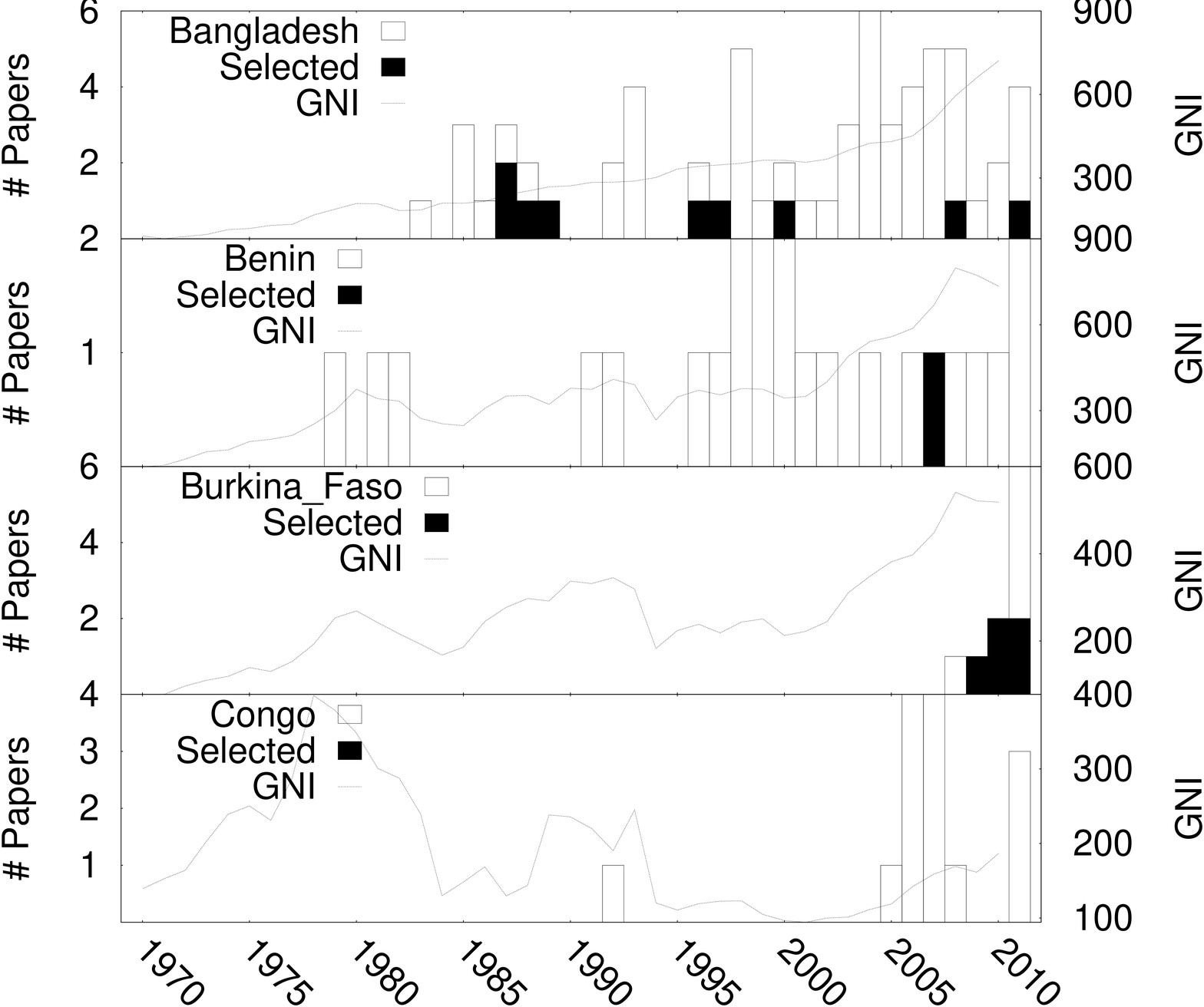}{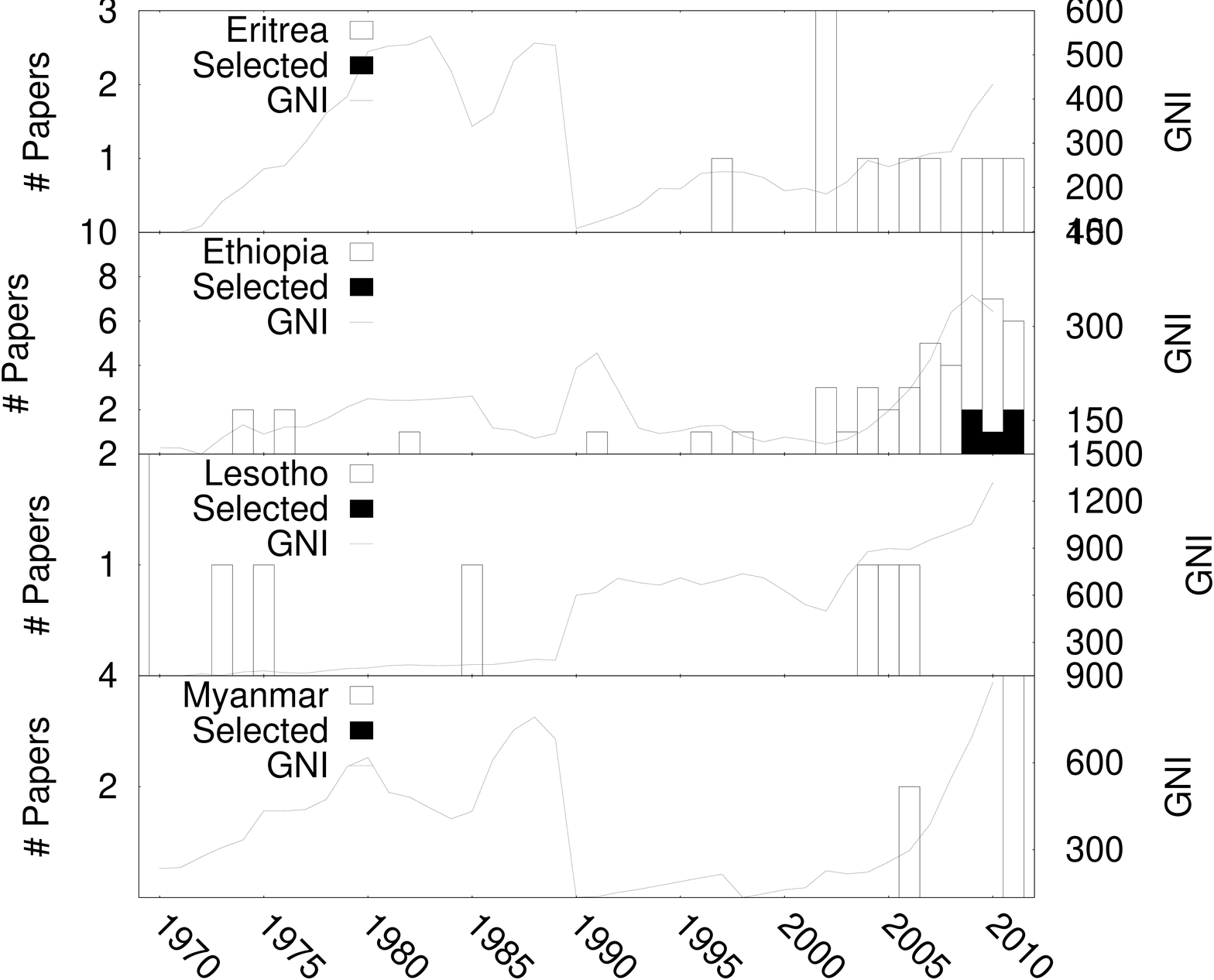}
\plottwo{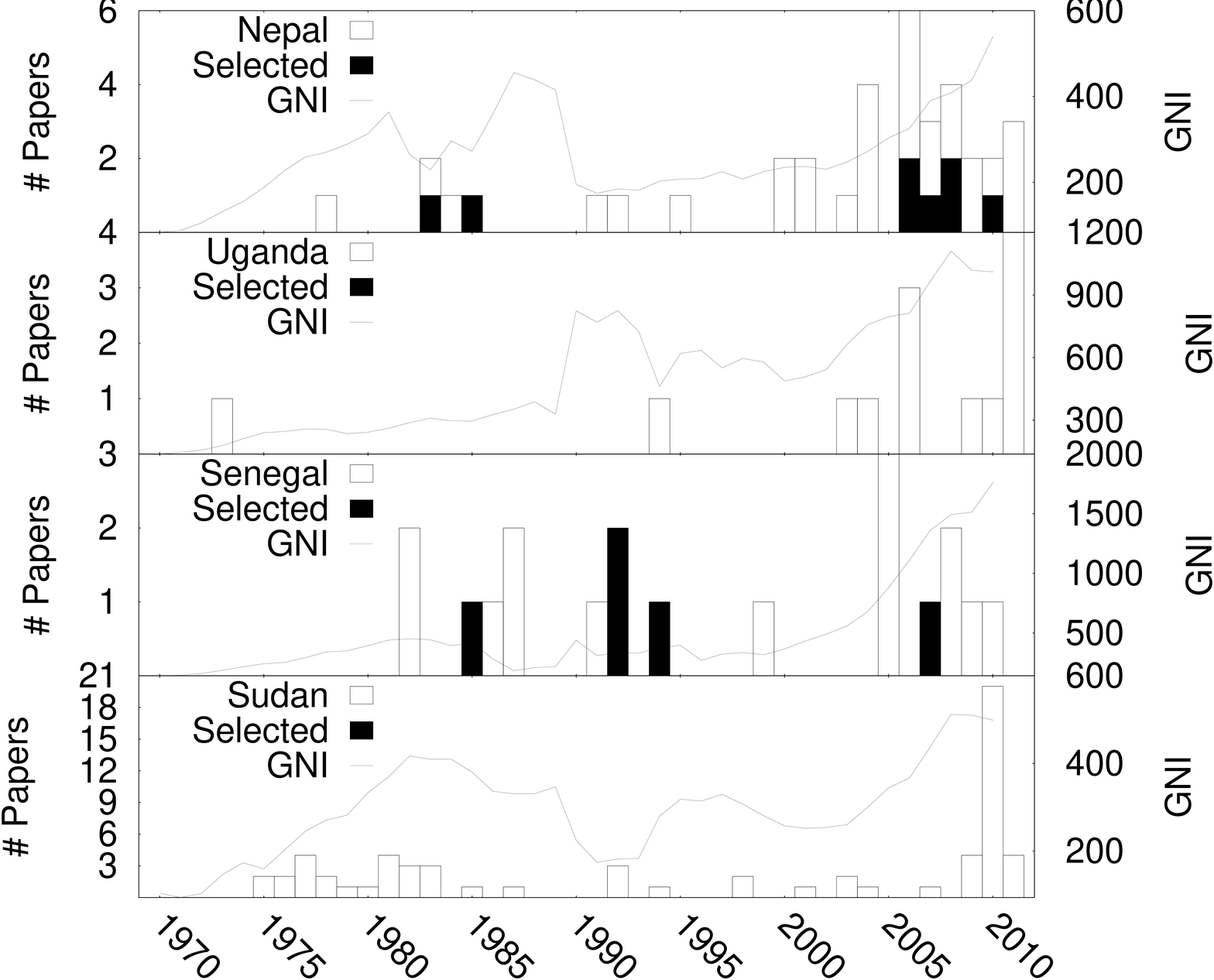}{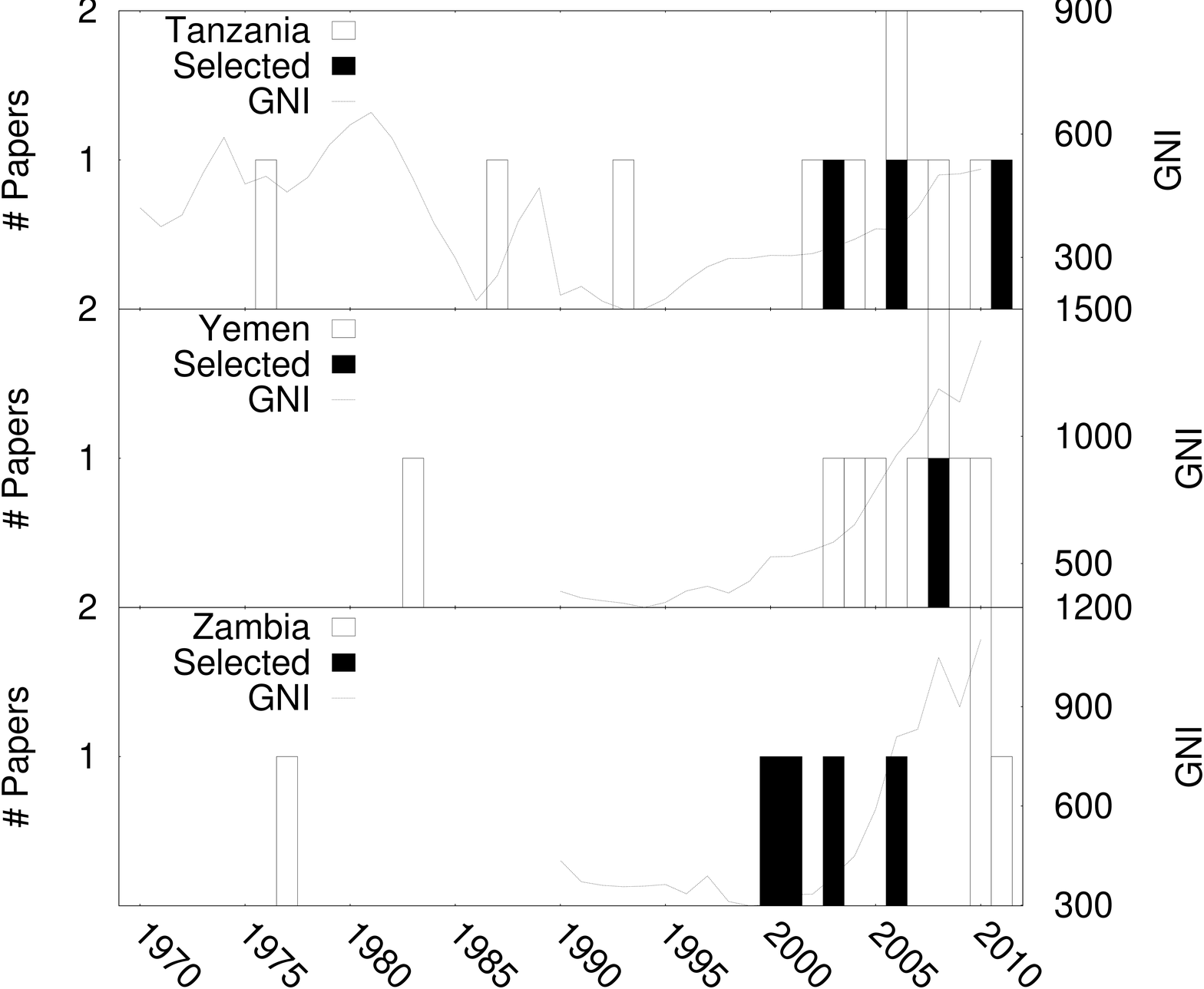}
\caption{As Fig.~\ref{fig1} but for the LDC.}
\label{fig6}
\end{figure}

Figures~\ref{fig1} to \ref{fig6} show the results for the number of papers per year, as well as the GNI per country per year, divided into Africa, South and Latin America, Asia, Europe, Oceania and LDCs, respectively. The white histograms are all the results as queried on ADS while the black histograms are for the selected mainstream astronomy journals, as mentioned above. Only countries with paper counts greater than five, in total, are shown in the figures, while excluded countries are shown in Table~\ref{tb:exc} along with their total number of papers in brackets. Furthermore, only the time span from 1970 onwards is shown, as before this date the number of papers is generally very scarce papers and will not aid our discussions.

We only considered results from the Astronomy database within ADS which also retrieves articles from the field of Geosciences. The search did not include results from the Physics database query within the ADS, which would undoubtedly increase the number of publications for any given country.

Several descriptions can be made upon visual inspection of Figures~\ref{fig1}~--~\ref{fig6}, these fall into five general categories which may complement the phases described by \citet{NG09}, described below:
\begin{enumerate}[i)]
\item Countries with history of publishing research articles, both in astronomy and other sciences.
\item Countries with a history of publishing in refereed journals, not including astronomy.
\item Countries where majority of research output is in astronomy.
\item Countries with little history of research publishing.
\item Countries that are difficult to place in the categories above, either due too little information and/or recent publishing activity.
\end{enumerate}

\section{Discussion and Conclusions}\label{discussion}
Based on the general descriptions above, most of the countries in Europe have a good history of publishing both in astronomy and other sciences while in Africa it is difficult to say anything quantitatively except in the cases of Egypt, Namibia and South Africa. In terms of the LDCs, more investment should be made in to general sciences and the culture of publishing. The Asian continent presents some interesting results, with countries like China and India have good history of publishing, both in astronomy and other science, and emerging countries like Thailand and Uzbekistan publish very few papers in any of these fields. In South and Latin America many countries have a history of publishing in both astronomy and other sciences, for example Argentina, Brazil, Chile, Mexico, Peru and Venezuela. Emerging countries with potential for developing astronomy further, due an already existing culture of publishing, include, Colombia and Uruguay.

We believe that the most successful country in developing astronomy will be that which already has a culture of publishing. In a number of countries, there appears to be a correlation between the countries GNI and the number of published papers. This may be related to an overall investment in Science and Technology, via job creation and making a country attractive to foreign scientists who may bring their expertise.

To put this in to context of major projects such as the IAU Office of Astronomy for Development\footnote{\url{http://www.astronomyfordevelopment.org/} last accessed 2012 October 07}, we believe that for a country to be successful in developing astronomy, within the lifetime of the office, it should have a well-established publishing record (not necessarily in Astronomy) or invest in bringing in outside expertise which can play a leading role in implementing courses at the university to help push for more papers to be published. One immediate example is that of Burkina Faso, where the University of Ouagadougou partnered with the University of Montreal in 2006, to develop an astronomy degree and build an observatory \citep{CTK11}. Also a level of investment from the country in science and technology would also aid improving the culture of publishing and individuals to disseminate their research and think critically about others' research. Similarly, \citet{BGO12} has outlined Turkish astronomy output from 1980 -- 2010, with further information about their astronomy community, including the impact of their publications. However, examples where an ethical conundrum about acquiring foreign expertise can be interpreted as a means in exchange for academic prestige \citep[][see also the various comments about the article, online and in Science Magazine on 2012 March 02]{B11} tells us more about how impact factors guide general research foundations on funding an institution and/or individual. Indeed no metric is fool proof and important strides are being done by various groups$\footnote{\url{http://www.cwts.nl/} last accessed 2013 July 01}^,\footnote{\url{http://info.scival.com/} last accessed 2013 July 01}^,\footnote{\url{http://researchanalytics.thomsonreuters.com/incites/} last accessed 2013 July 01}$.

Education and Public Outreach (EPO) programs both global scales, such as the IYA2009, and locally can play a key role in the development of astronomy in a country. For example, Mozambique used the momentum of the IYA2009 to develop local EPO programs and a a launching platform to develop astronomy at university level \citep{RPB11}. Similarly with the future construction of the Square Kilometer Array, decided between Australia and New Zealand, and South Africa and its partner countries (Botswana, Ghana, Kenya, Madagascar, Mauritius, Mozambique, Namibia and Zambia), the African continent is gearing up for the construction of an African Very Long Baseline Interferometer\footnote{\url{http://www.aerap.org/africanradioastronomy.php?id=32} last accessed 2013 July 01}.

In future work we would like to quantify the role each country has played on a paper, as a means to determine ``leadership''. A first indication, as mentioned above, a few of the countries are not leading any projects. However, we should determine what leading really means. In the era or large project we find more and more papers being in alphabetical order while normally the first author is the person who has played a leading role in the research. For example, the Research Excellence Framework\footnote{\url{http://www.ref.ac.uk/} lass accessed 2013 August 11}, in the United Kingdom, request that in a paper with more than 10 authors, the author should explain what their contribution to the paper was regardless if they are the first author or not. While no justification is required if the paper has less than 10 authors.

We only concentrated on the number of published papers to identify the global level of Astronomy Development. However, the ADS system has a number of other outputs which may be used for various studies, for example, the number of citations, the number of authors and their collaborations \citep[e.g.][]{N01}. This may also be an interesting project to visualize research collaborations\footnote{\url{http://orbitingfrog.com/post/34755190022/mapping-collaboration-in-astronomy} last accessed 2012 November 02}. 

\acknowledgements
This research has made use of NASA's Astrophysics Data System. The authors would like to thank Alberto Accomazzi and Edwin Henneken for bringing to our attention the ADS Mighty Search. VARMR acknowledges the Royal Astronomical Society, UK, for various travel grants. The South African SKA Project is acknowledged for funding the postdoctoral fellowship position at the University of Cape Town. We would like to thank comments from the community who supported the paper, in particular, Mattia Fumagalli, George Miley and Robert Simpson.  We thank an anonymous referee for constructive comments on the original manuscript.

\appendix

\section{World Countries}\label{wc}
The countries listed below are those considered for this study. These decision were made based on a gross national income of less than 14,365 USD in 2010. Those in italic are the Least Developed Countries (see also Appendix \ref{ldc}):

\begin{description}
\item[Africa]
	Algeria, {\it Angola}, {\it Benin}, Botswana, {\it Burkina Faso}, {\it Burundi}, Cameroon, Cape Verde, {\it Central African Republic}, {\it Chad}, {\it Comoros}, Cote d'Ivoire, {\it Democratic Republic of the Congo}, {\it Djibouti}, Egypt, {\it Equatorial Guinea}, {\it Eritrea}, {\it Ethiopia}, Gabon, {\it Gambia}, Ghana, {\it Guinea}, {\it Guinea-Bissau}, Kenya, {\it Lesotho}, {\it Liberia}, Libyan Arab Jamahiriya, {\it Madagascar}, {\it Malawi}, {\it Mali}, {\it Mauritania}, Mauritius, Morocco, {\it Mozambique}, Namibia, {\it Niger}, Nigeria, {\it Rwanda}, {\it Uganda}, {\it S\~ao Tom\'e and Principe}, {\it Senegal}, Seychelles, {\it Sierra Leone}, {\it Somalia}, South Africa, {\it Sudan}, Swaziland, {\it Togo}, {\it United Republic of Tanzania}, Tunisia, {\it Zambia}, Zimbabwe

\item[Asia]
	{\it Afghanistan}, {\it Bangladesh}, {\it Bhutan}, {\it Cambodia}, China, Democratic People's Republic of Korea, India, Indonesia, Iran, Iraq, Jordan, Kazakhstan, {\it Kiribati}, Kyrgyzstan, {\it Lao People's Democratic Republic}, Lebanon, Malaysia, Maldives, Mongolia, {\it Myanmar}, {\it Nepal}, North Korea, Pakistan, Palestine, Philippines, {\it Samoa}, {\it Solomon Islands}, Sri Lanka, Syrian Arab Republic, Tajikistan, Thailand, {\it Timor-Leste}, Turkmenistan, {\it Tuvalu}, Uzbekistan, {\it Vanuatu}, Vietnam, {\it Yemen}

\item[Europe]
	Albania, Armenia, Azerbaijan, Belarus, Bulgaria, Estonia, Georgia, Hungary, Kosovo, Latvia, Lithuania, Moldova, Montenegro, Poland, Romania, Russian Federation, Ukraine, Serbia, Turkey

\item[Latin America]
	Anguilla, Antigua and Barbuda, Argentina, Barbados, Belize, Bolivia, Brazil, Chile, Colombia, Costa Rica, Cuba, Dominica, Dominican Republic, Ecuador, El Salvador, Grenada, Guatemala, Guyana, {\it Haiti}, Honduras, Jamaica, Mexico, Montserrat, Nicaragua, Panama, Paraguay, Peru, Saint Kitts and Nevis, Saint Lucia, Saint Vincent and the Grenadines, Suriname, Uruguay, Venezuela

\item[Oceania]
	Cook Islands, Fiji, Marshall Islands, Micronesia, Nauru, Palau, Papua New Guinea, Tonga

\end{description}

\section{Least Developed Countries}\label{ldc}
The concept of Least Developed Countries, represents the poorest and weakest segment of the international community\footnote{\url{http://www.unohrlls.org/en/ldc/164/} last accessed 2012 July 20}. The list includes 48 countries; 33 in Africa, 14 in Asia and the Pacific and 1 in Latin America. In 2003 the Economic and Social Council of the United Nations used the following three criteria for the identification of the LDCs, as proposed by the Committee for Development Policy (CDP):
	
\begin{itemize}
\item A low-income criterion, based on a three-year average estimate of the gross national income per capita based on the World Bank Atlas method (under 992~USD for inclusion, above 1,190~USD to be removed from the list);
\item A human resource weakness criterion, involving a composite Human Assets Index based on indicators of: (a) nutrition; (b) health; (c) education; and (d) adult literacy; and
\item An economic vulnerability criterion, involving a composite Economic Vulnerability Index based on indicators of: (a) the instability of agricultural production; (b) the instability of exports of goods and services; (c) the economic importance of non-traditional activities (share of manufacturing and modern services in Gross Domestic Product); (d) merchandise export concentration; and (e) the handicap of economic smallness (as measured through the population in logarithm); and the percentage of population displaced by natural disasters.
\end{itemize}

To be added to the list, a country must satisfy all three criteria. To qualify for graduation, a country must meet the thresholds for two of the three criteria in two consecutive triennial reviews by the CDP. In addition, since the fundamental meaning of the LDC category, i.e. the recognition of structural handicaps, excludes large economies, the population must not exceed 75 million.

\clearpage

\clearpage

\clearpage

\clearpage

\clearpage

\clearpage

\clearpage

\begin{table}
\caption{Countries with five or less publications over the timespan of our studies. The number in brackets is the number of publications.}
\begin{tabular}{lll}
Country & Country & Country \\
\hline\hline
Afghanistan (1) & Gambia (0) & Nauru (0) \\
Angola (2) & Georgia (0) & Niger (0) \\
Anguilla (0) & Grenada (2) & North Korea (3) \\
Antigua and Barbuda (0) & Guatemala (5) & Palau (1) \\
Belize (1) & Guinea (0) & Rwanda (1) \\
Bhutan (1) & Guinea-Bissau (0) & Saint Kitts and Nevis (0) \\
Burundi (2) & Guyana (1) & Saint Vicent and the Grenadines (0) \\
Cambodia (1) & Haiti (2) & Samoa (2) \\
Cape Verde (3) & Kiribati (0) & Sao Tome and Principe (0) \\ 
Central African Republic (2) & Kosovo (4) & Seychelles (1) \\
Chad (1) & Lao (0) & Sierra Leone (1) \\
Comoros (0) & Liberia (0) & Solomon Islands (3) \\
Cook Islands (1) & Madagascar (2) & Somalia (4) \\
Cote d'Ivoire (0) & Malawi (1) & Suriname (0) \\
Djibouti (3) & Maldives (0) & Timor-Leste (0) \\
Dominica (0) & Mali (0) & Togo (0) \\
Dominican Republic (2) & Marshal Islands (1) & Tonga (0) \\
El Salvador (4) & Mauritania (1) & Tuvalu (0) \\
Equatorial Guinea (0) & Micronesia (1) & Vanuatu (1) \\
Gabon (2) & Mozambique (3) & \\ [0.4px]
\hline
\end{tabular}
\label{tb:exc}
\end{table}


\begin{thebibliography}{11}
\expandafter\ifx\csname natexlab\endcsname\relax\def\natexlab#1{#1}\fi

\bibitem[{Abt(2007)}]{A07}
Abt, H. 2007, Scientometrics, 72, 105

\bibitem[{Astronomy \& Astrophysics Survey~Committee(2001)}]{N01b}
Astronomy \& Astrophysics Survey~Committee, Board on Physics \&~Astronomy, S.
  S. B. N. R.~C. 2001, Astronomy and Astrophysics in the New Millennium (The
  National Academies Press)

\bibitem[{Bhattacharjee(2011)}]{B11}
Bhattacharjee, Y. 2011, Science, 334, 1344

\bibitem[{{Bilir} {et~al.}(2012){Bilir}, {Gogus}, {Onal}, {Ozturkmen}, \&
  {Yontan}}]{BGO12}
{Bilir}, S., {Gogus}, E., {Onal}, O., {Ozturkmen}, N.~D., \& {Yontan}, T. 2012,
  ArXiv e-prints

\bibitem[{{Carignan} {et~al.}(2011){Carignan}, {Turbide}, \&
  {Koulidiati}}]{CTK11}
{Carignan}, C., {Turbide}, L., \& {Koulidiati}, J. 2011, in IAU Symposium, Vol.
  277, IAU Symposium, ed. C.~{Carignan}, F.~{Combes}, \& K.~C. {Freeman}, 220

\bibitem[{{Hearnshaw}(2007)}]{H07}
{Hearnshaw}, J. 2007, IAU Special Session, 5, 9

\bibitem[{{Henneken} {et~al.}(2009){Henneken}, {Kurtz}, {Accomazzi}, {Grant},
  {Thompson}, {Bohlen}, \& {Murray}}]{HKA09}
{Henneken}, E.~A., {Kurtz}, M.~J., {Accomazzi}, A., {et~al.} 2009, Journal of
  Infometrics, 3, 1

\bibitem[{{Naicker} \& {Govender}(2009)}]{NG09}
{Naicker}, L., \& {Govender}, K. 2009, Communication Astronomy with the Public
  Journal, 7, 14

\bibitem[{Newman(2001)}]{N01}
Newman, M. E.~J. 2001, Phys. Rev. E, 64, 016131

\bibitem[{{Ribeiro} {et~al.}(2011){Ribeiro}, {Paulo}, {Besteiro}, {Geraldes},
  {Maphossa}, {Nhanonbe}, \& {Uaissine}}]{RPB11}
{Ribeiro}, V.~A.~R.~M., {Paulo}, C.~M., {Besteiro}, A.~M.~A.~R., {et~al.} 2011,
  in IAU Symposium, Vol. 260, IAU Symposium, ed. D.~{Valls-Gabaud} \&
  A.~{Boksenberg}, 522

\bibitem[{{Russo} \& {Christensen}(2010)}]{RC10}
{Russo}, P., \& {Christensen}, L.~L., eds. 2010, International Year of
  Astronomy 2009 -- Final Report

\end{thebibliography}
\end{document}